\documentclass[pre,aps,floats,superscriptaddress,floatfix,twocolumn]{revtex4}
\usepackage{amssymb,amsmath}
\usepackage{amsmath,amssymb}
\usepackage{graphicx}
\usepackage{psfrag}
\usepackage{color}
\usepackage{dcolumn}
\usepackage{bm}
\usepackage[normalem]{ulem}

\def\beq{\begin{equation}}
\def\eeq{\end{equation}}
\def\bea{\begin{eqnarray}}
\def\eea{\end{eqnarray}}
\begin{document}
\title{Marching on a rugged landscape: universality in disordered asymmetric exclusion processes}
\author{Astik Haldar}\email{astik.haldar@gmail.com}
\affiliation{Condensed Matter Physics Division, Saha Institute of
Nuclear Physics, HBNI, Calcutta 700064, India}
\author{Abhik Basu}\email{abhik.123@gmail.com, abhik.basu@saha.ac.in}
\affiliation{Condensed Matter Physics Division, Saha Institute of
Nuclear Physics, HBNI, Calcutta 700064, India}

\date{\today}

\begin{abstract}
 We { develop the hydrodynamic theory for} number conserving asymmetric exclusion processes with short-range random quenched disordered hopping rates, { which is one-dimensional Kardar-Parisi-Zhang (KPZ) equation with quenched columnar disorder}. 
 We show that when the system is away from half-filling, the universal spatio-temporal scaling of the density fluctuations is indistinguishable from its pure counterpart, with the  model belonging to the one-dimensional Kardar-Parisi-Zhang universality class. In contrast, close to half-filling,   the quenched disorder is relevant, leading to a new universality class.  We physically argue that the irrelevance of the quenched disorder when away from half-filling is a consequence of the averaging of the disorder by the propagating density fluctuations in the system. In contrast, close to half-filling the density fluctuations are overdamped, and as a result, are strongly influenced by the quenched disorder.
 
\end{abstract}

\maketitle

\section{Introduction}\label{intro}
Studies on the large-scale, macroscopic effects of quenched disorder in statistical mechanics models and condensed matter systems have a long tradition. For systems with  quenched
disorders, the impurities are fixed in particular configurations and do not evolve in time,
and, as a result, the disorder configuration is not in thermodynamic equilibrium. 
The effects of quenched disorder on driven, nonequilibrium systems are much less understood in comparison with their equilibrium counterparts. This is a question that could be of significance in a number of physical systems, e.g., systems involving flows in random media~\cite{rand-media}. In the absence of any general theories for nonequilibrium systems, it is useful to construct and study simple nonequilibrium models with quench disorders that are easy to analyse and yet can capture some basic features of more complex physical systems. This should help us to characterise and delineate regimes with different macroscopic behaviour. 

 In this article, we  address the generic scaling properties of one-dimensional ($1d$) Kardar-Parisi-Zhang (KPZ) equation with random quenched columnar disorder having short range correlations within a dynamic renormalisation group (RG) framework. As a realisation of the system, we briefly allude to  quenched-disordered asymmetric exclusion processes (TASEP) with particle number conservation,  and discuss how the quenched disordered $1d$ KPZ equation emerges in the hydrodynamic limit, that we use for our work. 


{ TASEPs with quenched disorder, both with open boundary conditions and periodic, have been investigated. 
 In these studies, notable issues addressed include improved mean field theories for disordered TASEP~\cite{sethna,stinch2}, generic steady state currents and bulk densities~\cite{lakatos}, correlation effects and boundary induced phase transitions in the presence of quenched disorder~\cite{ebrahim,foulaa-2008}, effects of a single bottleneck~\cite{greulich-2008}, interplay between quenched disorder and particle nonconservation due to Langmuir kinetics~\cite{greulich-2009}, defect induced phase transitions~\cite{schmidt}, fluctuations of domain walls~\cite{stinch3} and boundary layer analysis of the phase diagram of disordered TASEPs~\cite{sutapa}.} In this work, we focus on the large scale fluctuation properties of periodic TASEPs with quenched disorder. Recently, a number of simulations have been performed to study this, which generally point to larger fluctuations and slower dynamics~\cite{mustansir-prl,stinchcombe}. Here, we systematically construct the hydrodynamic theory for it, that to our knowledge is absent till the date, and show that it is the $1d$ KPZ equation with columnar disorder.

 We focus {on the $1d$ KPZ equation with quenched columnar disorder. We start by showing how that emerges in the long wavelength limit from generic number-conserving asymmetric exclusion processes with short-ranged quenched disordered hopping rates.} We then  use the resulting hydrodynamic equation to elucidate the universal scaling that characterises the local number fluctuations in the long wavelength limit. 
{ The principal results from our theory are as follows. { (i)  From the symmetry stand point, the hydrodynamic equations that we construct   are fundamentally different from its pure limit, which is the $1d$ Kardar-Parisi-Zhang (KPZ) equation. Unlike the $1d$ KPZ equation, these hydrodynamic equations for the disordered problem in general does not have the Galilean invariance, nor it satisfies the Fluctuation-Dissipation-Theorem (FDT). (ii) However, we show that away from the half-filling the relevant scaling exponents belong to the  universality class, rendering disorder irrelevant, and restoring Galilean invariance and FDT as {\em emergent symmetries} in the long wavelength limit.  (iii) { On the other hand, sufficiently close to half-filling,   the universal scaling properties are affected by the quenched disorder, and new universal scaling behaviour emerges.} In this case, there is no Galilean invariance or FDT in the long wavelength limit. This establishes the new universality class, as distinct from the $1d$ KPZ universality class. In particular, in the lowest order renormalised perturbation theory, we find the roughness exponent $\chi_h=5/8\approx 0.625$ and the dynamic exponent $z=7/4= 1.75$, both of which are larger than their values for the $1d$ KPZ equation. }  Our hydrodynamic equations reveal that the statistical steady states away from half-filling are marked by the presence of underdamped propagating density waves that move across the system, which vanish close to half-filling. 
The presence of the travelling density waves away from half-filling essentially imply that the density fluctuations encounter only average effects of the quenched disorder, which makes the latter ineffective in so far as the universal scaling is concerned with the scaling behaviour belonging to the $1d$ KPZ universality class. Since these waves vanish close to half-filling, this averaging effect disappears and the quenched disorders become relevant, leading to emergence of scaling different from the $1d$ KPZ universality class near half-filling. Our detailed analysis in this work establishes these intuitive expectations.

Our hydrodynamic theory is successful in delineating the possible universality classes in number conserving TASEP with short range quenched disordered hopping rates. This theory provides a general framework to understand the numerical results of Refs.~\cite{mustansir-prl,stinchcombe}. Due to the generic nature of the hydrodynamic theory, it is applicable for all number conserving asymmetric exclusion processes with short range quenched disorder. This opens up the possibility of systematic constructions of newer agent-based models for further detailed quantitative studies of the problem.    

The rest of this article is organised in the following manner. We construct the model and set up the continuum hydrodynamic equations of motion in Sec.~\ref{model}. Then in Sec.~\ref{dens}, we analyse the scaling properties of the density fluctuations in the nonequilibrium steady states. In particular, we set up the hydrodynamic equations that we use in Sec.~\ref{hydro-eqs}. We first study the symmetric limit of the problem in Sec.~\ref{dens-sep}, and move on to  the general, asymmetric case analysed in Sec.~\ref{dens-asep}. We show that the scaling of the density fluctuations in the half-filled limit is different from when the system is away from half-filling. We finally summarise in Sec.~\ref{summ}. Some of the technical details are discussed in Appendix for interested readers.

\section{Model}\label{model}

{ We first establish the $1d$ KPZ equation with random quenched columnar disorder as the hydrodynamic equation for the number conserving quenched disordered TASEP, which was phenomenologically proposed in Ref.~\cite{mustansir-prl}.} We consider an asymmetric exclusion process (TASEP) on a closed ring with $L$ sites.  The unidirectional rate of hopping $m_i,\,(1\leq i\leq L)$ from site $i$ to $i+1$ is a positive 
definite time-independent random number, i.e., $m_i$ is {\em quenched}; see Fig.~\ref{model-tasep}.

\begin{figure}[htb]
\includegraphics[width=7cm]{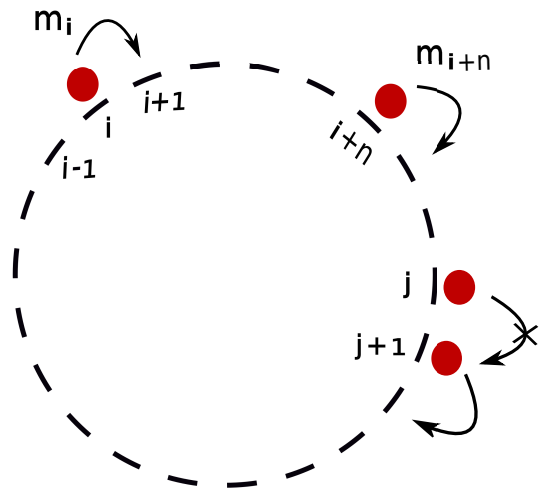}
 \caption{Schematic model diagram for asymmetric exclusion process on a ring. Broken lines forming a circular loop represent the lattice sites; small filled red circles refer to the particles which can hop only in one direction, subject to exclusion. Parameters $m_i$ and $m_{i+n}$ represent 
hopping rates at sites $i$ and $i+n$; $m_i\neq m_{i+n}$ in general (see text for more details).} \label{model-tasep}
\end{figure}

   The quenched disordered hopping rates make the system inhomogeneous and break the translational invariance along the ring. The occupation $n_i$ for site $i$, can be either 0 or 1 and follows
\begin{equation}
 \frac{\partial n_i(t)}{\partial t}= m_{i-1} n_{i-1} (1-n_{i}) - m_{i} n_{i} (1-n_{i+1}),
\end{equation}
together with $0<m_i<1$.

In the thermodynamic limit with $L\rightarrow \infty$, it is  
convenient to introduce a quasi-continuous coordinate $x=i/L$, such that { $0\leq 
x\leq 1$} with $L\rightarrow \infty$ and $\rho(x,t)\equiv n_i(t),\,m(x)=m_i$. In this continuum 
limit using a gradient expansion up to the second order in spatial gradients, we 
obtain  
\begin{eqnarray}
 \frac{\partial \rho(x,t)}{\partial t}&=& \nu\frac{\partial^2}{\partial x^2}\rho(x,t)-\frac{\partial}{\partial x} [m(x)\rho(x,t) 
(1-\rho(x,t))] \nonumber \\ &+& \frac{\partial f(x,t)}{\partial x},\label{fulrho}
\end{eqnarray}
{ where we have neglected terms that are ${\cal O}(1/L)$ or higher, and are subleading (in a scaling sense) in gradient-expansions or nonlinearities}; $f$ is a white noise that models the stochasticity of  the 
underlying microscopic dynamics, and $\nu$ is diffusion coefficient: $\nu\sim 
m_0/L$ { where $m_0$ is the mean value of $m(x)$ (see below);  see Appendix \ref{derivation};  see also Refs.~\cite{kiran,erwin-lk} for discussions on the principles to obtain the hydrodynamic limit for the Burgers equation}.  We choose $f(x,t)$ to a zero-mean, Gaussian distributed noise with a 
variance
\begin{equation}
 \langle f(x,t) f(0,0)\rangle = 2D\delta(x)\delta (t),\label{noise-vari}
\end{equation}
where $D>0$ is the nonequilibrium analogue of temperature.
We further write
$m(x)=m_0+\delta m(x)$, where $m_0=\overline{m(x)}$, the mean of $m(x)$ as defined above, and $\delta m(x)$ is the local (in space) fluctuation of $m(x)$ about $m_0$; $\overline{ \delta m(x)} =0$~\cite{foot1}. We assume 
$\delta m(x)$ to be Gaussian-distributed with a variance
\begin{equation}
 \overline{ \delta m(x)\delta m(0)} = 2\tilde D\delta (x).\label{mvari}
\end{equation}
Equivalently, in the Fourier space,
\begin{equation}
 \overline {\delta m(k_1,\omega_1)\delta m(k_2,\omega_2)}=2\tilde D\delta (k_1+k_2)\delta(\omega_1 +\omega_2)\delta(\omega_1).\label{mvariF}
\end{equation}
The last $\delta$-function factor (\ref{mvariF}) appears due to the fact $\delta m(x)$ is time-independent, and so is the correlator on the rhs of (\ref{mvari}).
Thus the quenched disorder is short ranged. Here,  $\langle 
...\rangle$ implies averages over the annealed (time-dependent) noise distribution, where 
as an ``overline'' refers to averages done over the random quenched disorder 
distribution. Our aim here is to analyse the spatio-temporal scaling properties of (\ref{fulrho}) in the hydrodynamic limit.


\section{Scaling of the density fluctuations}\label{dens}

\subsection{Hydrodynamic equations and universal scaling exponents}\label{hydro-eqs}

In order to extract the scaling behaviour, we must solve for $\rho(x,t)$ from (\ref{fulrho}) as a function of $f(x,t)$ and $\delta m(x)$. It is convenient to express $\rho(x,t)$ as
 $\rho(x,t)=\rho_0+\phi(x,t)$, where { $\int_0^1 dx 
\rho(x,t)=\rho_0 $} and $\int dx \phi (x,t)=0$  at all time $t$. 
Fluctuation $\phi(x,t)$ then satisfies
{
\begin{eqnarray}
 \frac{\partial \phi(x,t)}{\partial t}&=& \nu \frac{\partial^2 \phi(x,t)}{\partial x^2} +  \lambda_1 \frac{\partial \phi(x,t)}{\partial x}+ \lambda_m \frac{\partial \delta m (x)}{\partial x}  \nonumber \\ &&    - \frac{\lambda}{2} \frac{\partial  \phi^2(x,t)}{\partial x} + \lambda_2 \frac{\partial}{\partial x}[\delta m(x) \phi(x,t)]  \nonumber\\&&  - \lambda_3 \frac{\partial}{\partial x}[\delta m (x) \phi^2(x,t)] + \frac{\partial f(x,t)}{\partial x},
\end{eqnarray}
where   $\lambda=2m_0/L$, $\lambda_m=(\rho_0^2-\rho_0)/L$, $\lambda_1=m_0(2\rho_0-1)/L$, $\lambda_2=(2\rho_0-1)/L$, $\lambda_3=1/L$. Notice that  
$\phi(x,t)$ is driven by {\em two} conserved noises - quenched noise 
$\partial_x\delta m (x)$ and annealed noise $\partial_x f(x,t)$.

To proceed further, we decompose $\phi(x,t)$ into two parts:  $\phi(x,t)=\psi(x) 
+ \delta\rho (x,t)$; $\int dx \psi(x) =0 =\int dx \delta\rho(x,t)$. Time-independent function $\psi(x)$ satisfies 
 \begin{eqnarray}
 &&-\nu_{\psi} \partial_{xx} \psi - \lambda_{1\psi} \partial_x \psi -  
\lambda_m \partial_x \delta m + \frac{\lambda_{\psi}}{2} \partial_x [\psi^2] 
\nonumber \\&&- \lambda_{2\psi}\partial_x[\delta m \psi ] + 
\lambda_{3\psi} \partial_x[\delta m \psi^2] = 0. \label{psieq}
\end{eqnarray}

In contrast, $\delta \rho (x,t)$ satisfies the time-dependent equation
\begin{eqnarray}
 \frac{\partial\delta \rho}{\partial t} &=& \nu_{\rho} \partial_{xx} 
\delta\rho + \lambda_{1\rho} \partial_x \delta\rho - 
\frac{\lambda_{\rho}}{2} \partial_x \delta\rho^2 + \lambda_{2\rho}\partial_x[\delta m \delta\rho]   \nonumber \\&-&  \lambda_{\rho\psi}\partial_x[\delta\rho \psi] - \lambda_{3\rho} \partial_x[\delta m \delta\rho^2] - \lambda_{3\rho\psi}\partial_x[\delta m \delta\rho \psi] \nonumber\\ &+& \partial_xf.\label{eqdelrho}
\end{eqnarray}
{Here the model parameters  $\nu_\psi$, $\nu_\rho$ are proportional to $\nu$; $\lambda_\psi$, $\lambda_{\rho}$, $\lambda_{\rho\psi}$ are proportional to $\lambda$;  $\lambda_{1\psi}$, $\lambda_{1\rho}$ are proportional to $\lambda_1$; $\lambda_{2\psi}$, $\lambda_{2\rho}$ are proportional to $\lambda_2$ and $\lambda_{3\psi} $, $\lambda_{3\rho}$, $\lambda_{3\rho\psi}$ are proportional to $\lambda_3$. } We do not use the same set of model parameters in (\ref{psieq}) and (\ref{eqdelrho}) in order to allow for different scaling by $\psi$ and $\delta\rho$ and hence by the associated model parameters in the long wavelength limit.
Clearly, in the absence of any quenched disorder ($\delta m=0$), $\psi(x)=0$, and $\delta\rho(x,t)$ becomes the local fluctuations of the density around its mean $\rho_0$, and satisfies the $1d$ Burger's equation~(\ref{burg1}) (see below). Notice that $\psi(x)$ that is frozen in time is entirely driven by the quenched noise $\delta m(x)$. On the other hand, $\delta\rho(x,t)$ is driven by a time-dependent (annealed) noise $\partial_x f$ additively (i.e., as a ``source'') and also by the quenched noise $\delta m(x)$ multiplicatively. In addition, $\psi(x)$ enters into the dynamics of $\delta\rho(x,t)$, where as $\psi(x)$ itself is independent of $\delta\rho(x,t)$. 

Equations~(\ref{psieq}) and (\ref{eqdelrho}) have two linear terms with first order spatial gradients having coefficients proportional to $\lambda_1$: For example, in (\ref{eqdelrho}) the term $\lambda_{1\rho} \partial_x \delta\rho$ implies the existence of underdamped propagating modes for $\delta\rho$. This linear propagating mode cannot be removed by going to the co-moving frame, i.e., the model does not admit Galilean invariance. This is a consequence of the fact that the variance (\ref{mvari}) manifestly breaks Galilean invariance and the explicit appearance of $\delta m$ and $\psi$ in (\ref{eqdelrho}). Interestingly, in the half-filled limit, both these linear terms in (\ref{psieq}) and (\ref{eqdelrho}) vanish, since $\lambda_1=am_0(2\rho_0-1)$ vanishes. Thus for $\rho_0\approx 1/2$, the dynamics of $\delta\rho(x,t)$ has no underdamped propagating modes. This has a significant bearing on what follows below.

With the transformation $\delta\rho(x,t)=\partial_x h(x,t)$, Eq.~(\ref{eqdelrho}) reduces to
\begin{eqnarray}
  \frac{\partial h(x,t)}{\partial t} &=& \nu_{\rho} \partial_{xx} 
h + \lambda_{1\rho} \partial_x h - 
\frac{\lambda_{\rho}}{2}  (\partial_x h)^2  \nonumber \\&+&
\lambda_{2\rho} \delta m \partial_x h 
-  \lambda_{\rho\psi}(\partial_x h) \psi
 \nonumber \\&-& \lambda_{3\rho} \delta m (\partial_x h)^2 -
\lambda_{3\rho\psi}\delta m \partial_x h \psi + f.\label{eqh}
\end{eqnarray}
Equation~(\ref{eqh}) represents a growing surface in the presence of quenched columnar disorders; see also Ref.~\cite{mustansir-prl,stinchcombe}. In the pure limit, i.e., $\psi(x)=0$, (\ref{eqh}) reduces to the standard KPZ equation. { Just as Eq.~(\ref{eqdelrho}) has propagating modes, Eq.~(\ref{eqh}) too has propagating modes that vanish near half-filling.} Further, just as Eq.~(\ref{eqdelrho}) is not invariant under a Galilean transformation, Eq.~(\ref{eqh}) is not invariant under an equivalent tilt of the surface. The presence of the columnar disorders manifestly breaks the tilt invariance of the ordinary KPZ equation. 

{ Noting that $\psi(x)$ is independent of $t$, we define the universal scaling exponents that characterise the auto-correlation functions of $\psi(x)$ and $h(x,t)$:
\begin{eqnarray}
 &&C_{\psi\psi}(x)\equiv \overline{ \psi(x)\psi(0)}= |x|^{2\chi_\psi}\\
 &&C_{\rho\rho}((x,t)\equiv \langle \delta \rho(x,t)\delta\rho(0,0\rangle = |x|^{2\chi_\rho}\varTheta_\rho(|x|^z/t),\\
 && C_{hh}(x,t)\equiv \overline{\langle [h(x,t) - h(0,0)]^2\rangle}= |x|^{2\chi_h}\varTheta(|x|^z/t),
\end{eqnarray}
where $\chi_h=1+\chi_\rho$. 
Scaling exponents $\chi_\psi,\, \chi_h$ are the roughness exponents of $\psi$ and 
 $h$ respectively; $z$ is the dynamic exponent of $\delta\rho$. }

 \subsection{Universality in closed disordered TASEP: review of numerical studies}
 
  Before we proceed with our hydrodynamic equations, let us review the recent numerical results on number conserving quenched disordered TASEP.  Ref.~\cite{mustansir-prl} studied this problem with a binary distribution for $m_i$, that is controlled by two parameters, one describing the ratio of the strengths of the {\em strong} bonds and {\em weak} bonds, and the other representing the fraction of weak bonds. Results in Ref.~\cite{mustansir-prl} broadly reveal that (i) away from the half-filling, the quenched disorder is ineffective, having no effect on the large scale, macroscopic scaling properties, with  the system belonging to the well-known $1d$ KPZ universality class, (ii) in contrast, close to the half-filling, the quenched disorder affects the large scale scaling properties, with the emergence of a {\em new universality class}. In particular, close to half-filling Ref.~\cite{mustansir-prl} reported scaling exponent $\beta=\chi_h/z\approx 0.42 >0.33$, its value away from the half filling, which is also its value for the pure $1d$ KPZ equation. In a subsequent numerical study on the same model at half filling, Ref.~\cite{stinchcombe} found the dynamics at half-filling to be distinctly slower than for the pure model. In particular, Ref.~\cite{stinchcombe} found $\chi_h\approx 1.05$ and $z\approx 1.7$ (corresponding to $\beta\approx 0.62$), both being higher than their values for the pure $1d$ KPZ problem. While the precise values of the scaling exponents in Ref.~\cite{mustansir-prl} and Ref.~\cite{stinchcombe} do not agree quantitatively, they both indicate the existence of a new universality class near the half filling. We intend to complement these numerical results by studying a hydrodynamic theory that we construct here. However, unlike the studies in Refs.~\cite{mustansir-prl,stinchcombe}, we use a short-ranged Gaussian distributed quenched disorder. We are not aware of any numerical study  analogous to those in Refs.~\cite{mustansir-prl,stinchcombe}, but with Gaussian distributed quenched disorder. Nonetheless, due to the short range nature of the disorder considered here, we expect our model should display the same universal behavior as those in Refs.~\cite{mustansir-prl,stinchcombe}.

 \subsection{Underdamped propagating waves and the universality classes}

  We now revisit and expand our heuristic arguments in Sec.~\ref{intro} above that close to half-filling, emergence of a different universality class is expected. Particle-hole symmetry of the model implies that at $\rho_0=1/2$, the system has equal number of particles and holes, where as for $\rho_0 > (<) 1/2$, the system has more particles (holes) than holes (particles). We further note that for $\rho_0 > (<) 1/2$, the system admits underdamped propagating density waves with speeds proportional to the total excess particles (holes) in the system with respect to the half-filled limit; see Eq.~(\ref{eqdelrho}). Thus local density fluctuations of particles (or holes) move across the system as propagating waves without significant damping, thereby encountering only an average effect of the quenched hopping rates. This renders the quenched hopping rates ineffective.  On the other hand, close to the half-filled limit,  the propagating modes are absent, and the density fluctuations are overdamped and can relax only by diffusively exchanging particles with the neighbouring regions with excess or deficit particles. To understand its consequence, let us imagine the system to be made up of small regions, characterised by given hopping rates different from the neighbouring patches. Due to the heterogeneity of the hopping rates, particles tend to accumulate behind a patch of lower hopping rates in the statistical steady states. Thus, as a result of the frozen unequal hopping rates varying from regions to regions and in the absence of any propagating modes, local regions behind lower hopping rates on average would typically have more particles than regions behind higher hopping rates in the statistical steady states. 
This evidently makes the density correlation function less smooth than that for the pure system, corresponding to a larger $\chi_\rho$ (and hence a larger $\chi_h$) here close to half-filling. Our detailed calculations below confirm this intuitive physical picture.

\section{Symmetric exclusion process with quenched disorder}\label{dens-sep}

Before we attempt to solve (\ref{psieq}) and (\ref{eqdelrho}) or (\ref{eqh}),
it is instructive to first look at the symmetric limit of the problem - number conserving {\em symmetric} exclusion processes (SEP) with random quenched disordered hopping rates. In SEP, particle movement is bidirectional, subject to exclusion; see Fig.~\ref{model-sep}.

\begin{figure}[htb]
\includegraphics[width=7cm]{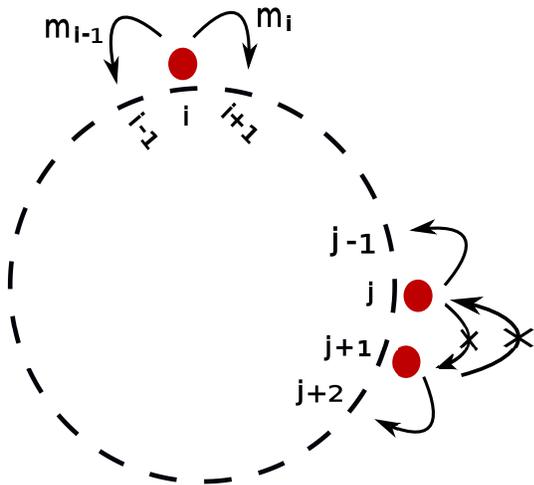}
 \caption{Schematic model diagram for symmetric exclusion process on a ring. Broken lines (forming a circular loop) represent lattice sites. Small filled red circles refer to the particles which can hop in either direction, subject to exclusion (see text for more details).}\label{model-sep}
\end{figure}

Again writing the local fluctuating density as the sum of $\psi(x)$ and $\delta\rho(x,t)$, the effective long wavelength equations for $\psi(x)$ and $\delta\rho(x,t)$ after discarding the irrelevant terms (in a scaling sense) read
\begin{eqnarray}
 &&D_sm_0\partial_{xx}\psi(x) = 0,\label{sep-psi}\\
 &&\frac{\partial\delta\rho(x,t)}{\partial t}-m_0\partial_{xx}\delta\rho = \partial_x f, \label{sep-rho}
\end{eqnarray}
to the leading order in fluctuations;
see also Appendix \ref{derivation}. We thus find that $\psi(x)$ and $\delta\rho(x,t)$ are mutually decoupled to the leading order in fluctuations at all $\rho_0$, and the dynamics of $\delta\rho$ is  insensitive to $\rho_0$. With $\delta\rho = \partial_x h$, (\ref{sep-rho}) reduces to the well-known Edward-Wilkinson model for growing surfaces driven by a white noise~\cite{stanley}:
\begin{equation}
 \frac{\partial h}{\partial t}-\nu_\rho\partial_{xx}h = f.\label{ew}
\end{equation}
Equation~(\ref{sep-psi}) implies $\psi=0$, where as due to the linearity of (\ref{sep-rho}) or (\ref{ew}), the corresponding scaling exponents are known {\em exactly}: $\chi_h=1/2,\,z=2$. Thus,
\begin{equation}
 C_{hh}(r,t)\equiv\overline{\langle [h(x+r,t)-h(x,0)]^2\rangle }=r\varTheta(r^2/t).
\end{equation}
 Since $\psi(x)=0$, the steady state itself is uniform when viewed at sufficiently large scales. Further, with the scaling of $C_{hh}$ being indistinguishable from its pure counterpart, we conclude that any coarse-grained measurements of the density fluctuations cannot detect existence of any short range quenched disorder in a periodic SEP.

\section{Scaling in the asymmetric case}\label{dens-asep}



\subsection{Linear theory}

Having discussed the simpler quenched disordered SEP with particle number conservation, we now go over to the more general (and more complex as we will see below)  corresponding asymmetric case.

It is useful to first study the linearised version of the asymmetric case. 
Equation~(\ref{psieq}), when away from the half-filled limit, $\psi(k)= -(\lambda_m/\lambda_{1\psi})\delta m(k)$ in the Fourier space in the long wavelength limit, meaning $\psi(x)$ is just as rough as $\delta m(x)$. In this linear limit, this implies $\psi(x)$ can be obtained by minimising an effective free energy ${\cal F}_1:\;\delta {\cal F}_1/{\delta \psi}=0$, where
\begin{equation}
 {\cal F}_1=\int dx \left[\lambda_{1\psi}\psi(x)^2/2 + \lambda_m\psi(x)\delta m(x) \right].
\end{equation}
 On the other hand, close to the half-filled limit, $\psi(k)=i\lambda_m/(k\nu_\psi) \delta m(k)$, implying that $\psi(x)$ is distinctly rougher than $\delta m(x)$ sufficiently close to the half-filled limit. Again, in this linear limit, $\psi(x)$ can be obtained by minimising an effective free energy ${\cal F}_2$: $\delta {\cal F}_2/{\delta \psi}=0$, where 
\begin{equation}
{\cal F}_2=\int dx \left[\nu_\psi (\partial_x\psi)^2/2 - \lambda_m\psi \partial_x \delta m \right].
\end{equation}

{

Away from the half-filled limit, the equal-time auto-correlation functions of $\psi$ and $\delta\rho$  are given by
\begin{eqnarray}
 &&\overline {|\psi(k)|^2}=\frac{2\tilde D\lambda_m^2}{\lambda_{1\psi}^2},\label{lin-corr-psi-nothalf}\\
 &&\overline{\langle |\delta\rho(k,t)|^2\rangle}=\frac{D}{\nu_\rho},\label{lin-corr-rho-nothalf}\\
 &&\overline{\langle |h(k,t)|^2\rangle}=\frac{D}{\nu_\rho k^2};\label{lin-corr-h-nothalf}
\end{eqnarray}
see also Appendix \ref{functional}.  Clearly, $\psi(k)$ and $\delta\rho(k,t)$, and hence $\psi(x)$ and $\delta\rho(x,t)$ scale the same way. This no longer holds true near the half-filled limit, for which the equal-time correlators are
%
\begin{eqnarray}
 &&\overline {|\psi(k)|^2}=\frac{2\tilde D\lambda_m^2}{\nu^2_\psi k^2},\label{rcorr-psi-half} \\
 &&\overline{\langle |\delta\rho(k,t)|^2\rangle}=\frac{D}{\nu_\rho},\label{rcorr-rho-half}\\
 &&\overline{\langle |h(k,t)|^2\rangle}=\frac{D}{\nu_\rho k^2 }.\label{rcorr-h-half}
\end{eqnarray}
Thus, $\psi(k)$ scales differently from $\delta\rho(k,t)$. In particular, 
$\psi(k)$ is {\em more relevant} (in a scaling sense) than $\delta\rho(k)$ in the infra-red (long wavelength) limit. It further implies that $\psi(x)$ is more relevant than $\delta\rho(x)$ (in a scaling sense). Interestingly, in the linear theory, $\overline{\langle |\delta\rho(k,t)|^2\rangle}$ and $\overline{\langle |h(k,t)|^2\rangle}$ do not depend upon the filling-factor.

 In the linear theory, 
 the exponents are known {\em exactly}. For instance, for $\rho_0\neq 1/2$, $\chi_\psi=-1/2$, since $\overline{ \psi(x)\psi(0)}$ is proportional to $\delta (x)$ that indeed scales as $1/|x|$; see Eqs.~(\ref{psi-nothalf-corr}) { below}. Similarly, $\chi_h=1/2$ and $\chi_\rho=-1/2$. Further, linearity of the $\delta \rho(x,t)$-dynamics implies $z=2$. On the other hand when $\rho_0\approx 1/2$, Eqs.~(\ref{psi-half-corr}) and (\ref{h-half-corr}) give $\chi_\psi=1/2=\chi_h$ and $\chi_\rho=-1/2$; $z$ continues to remain 2 at the linear level. It remains to be seen how the various nonlinear terms may affect these scaling exponents.
 }

 \subsection{Anharmonic effects}

Nonlinear terms and the propagating modes in (\ref{psieq}) and (\ref{eqdelrho}) may change the above simple picture in terms of ${\cal F}_1$ and ${\cal F}_2$ and the corresponding scaling exponents in the linear theory, that we seek to find now. This cannot be done exactly due to the nonlinear terms. Furthermore, 
na\"ive perturbative theory can produce diverging corrections to the model 
parameters in (\ref{psieq}), as happens for ordinary Burgers 
equation~\cite{stanley,fns}.  This calls for more systematic  and refined treatments.

{  At this point, we note that in the approach of Refs.~\cite{spohn1,spohn2,spohn3,spohn4}, by using a combination of one-loop approximation and numerical solutions of the mode-coupling equations for the correlation functions, detailed predictions for the correlation functions of the conserved variables in the $1d$ Burgers equation are made. 
The present model, however, has additional complications due to the possibility of diverging vertex corrections (see below). How the nonlinear fluctuating hydrodynamics approaches developed in Refs.~\cite{spohn1,spohn2,spohn3,spohn4} is still not settled. We here instead use the dynamic RG 
method.} 
{ Dynamic RG methods applied on continuum driven hydrodynamic models have a long history of studies in statistical mechanics. These are well-established methods, particularly suitable to delineate the universality classes of the driven systems; see Refs.~\cite{fns,kpz,erwin-epl,natter,janssen1,toner-tu,abhik_berlin,uwe} for  applications of Dynamic RG methods in related models.}

{ While the Dynamic RG method is well-documented~\cite{halpin} in the literature, we give a brief outline of the method for the convenience of the reader. The momentum shell Dynamic RG procedure consists of integrating over the short wavelength 
Fourier modes
of $\psi(x)$ and $\delta\rho(x,t)$, followed by a rescaling of lengths and time.
More precisely, we  follow the standard approach of initially restricting wavevectors  
to lie in a 1D Brillouin zone: $|{ q}|<\Lambda$, where 
$\Lambda$ is an
ultra-violet cutoff,  presumably of order the inverse of the lattice spacing $a$, although its precise value has no effect on our  results. The density fields ${
\psi}(x)$ and $\delta\rho(x,t)$
are separated into high and low wave vector parts
$\psi(x)=\psi^>(x)+\psi^<(x)$ and $\delta\rho(x,t)=\delta\rho^>(x,t) + \delta\rho^<(x,t)$,
where $\psi^>(x)$ and $\delta\rho^>(x,t)$ have support in the large wave vector  (short wavelength) 
range $\Lambda
e^{-\delta\ell}<| q|<\Lambda$, while $\psi^<(x)$ and $\delta\rho^<(x,t)$ have support in the small 
wave vector (long wavelength) range $|{ q}|<e^{-\delta\ell}\Lambda$.
We then integrate out $\psi^>(x)$ and $\delta\rho^>(x,t)$. This integration is done perturbatively 
in 
 the anharmonic couplings in (\ref{action1}); as usual, this perturbation theory 
can be represented by Feynman graphs, with the order of perturbation theory 
reflected by the number of loops in the graphs we consider. After this 
perturbative step, we rescale lengths, 
with ${ x}={ x}' e^{\delta\ell}$,  so as to restore the UV cutoff back to 
$\Lambda$ and {also time by $t=t'e^{z\delta\ell}$, where $z$ is the dynamic exponent}.  This is then followed
by rescaling the long wave length part of the fields; see Appendix \ref{not half-filled}.}

 It is important to note that there are two classes of Feynman diagrams: (i) the 
first kind survives in the limit of vanishing disorders and originate from the standard
Burgers-nonlinear term $\lambda_{\rho}\partial_x \delta\rho^2$ in (\ref{eqdelrho}), or the $\lambda_{\rho} dx dt \hat\rho \partial_x \delta\rho^2$-term in the action functional 
(\ref{action1}); (ii) the second type originates from the nonlinear terms 
 that involve the disorder in (\ref{psieq}) 
and (\ref{eqdelrho}), or in the  action functional (\ref{action1}); see 
Appendix \ref{functional}. 

We distinguish two cases: (i) away from half-filling ($\rho_0\neq 1/2$) and (ii)  
close to half-filling $\rho_0\approx 1/2$.

\subsection{Pure periodic asymmetric exclusion process}

In the absence of any quenched disorder, $\delta m(x)=0$ and the hopping rate is uniform everywhere. This gives $\psi(x)=0$ for all $x$ identically, and Eq.~(\ref{eqdelrho}) reduces to
\begin{equation}
 \frac{\partial\delta\rho}{\delta t}=\nu_{\rho}\partial_{xx}\delta\rho  - \frac{\lambda_{\rho}}{2}\partial_x\delta\rho^2 +\partial_x f,\label{burg1}
\end{equation}
where the $\lambda_{1\rho}\partial_x\delta\rho$-term in (\ref{burg1}) has been removed by using the Galilean invariance of (\ref{burg1}); $\nu_\rho$ is the diffusivity.  Equation~(\ref{burg1}) implies a current density $J$ given by
\begin{equation}
 J=-\nu_{\rho}\partial_x \delta\rho + \frac{\lambda_{\rho}}{2} \delta\rho^2- f.\label{burg-curr}
\end{equation}
With $\delta\rho=\partial_x h$, $h$ satisfies the 1D KPZ equation driven by a white noise:
\begin{equation}
 \frac{\partial h}{\partial t}=\nu_{\rho}\partial_{xx} h  - \frac{\lambda_{\rho}}{2}(\partial_x h)^2 + f.\label{kpz1}
\end{equation}
 The scaling exponents of (\ref{kpz1}) are {\em exactly} known  due to the Galilean invariance of (\ref{burg1}) or (\ref{kpz1}) together with an FDT, with $\chi_h=1/2,\,z=3/2$ with an exact relation $\chi_h+z=2$~\cite{stanley,fns}. This corresponds to $\chi_\rho = -1/2$.

\subsection{Away from half-filling}

When $\rho_0\neq 1/2$, the Feynman diagrams of the second type defined above are all  
finite. In contrast, the Feynman diagrams of the first type remain infra-red 
divergent as they are for the pure KPZ/Burgers problem in $1d$.  At a more technical level, the dimensionless coupling constants that depend upon the disorder variance $\tilde D$ and appear in the diagrammatic expansions of the model parameters when away from half-filling, are all {\em irrelevant} (in the RG sense) near the Gaussian fixed point (see Appendix \ref{appen-gauss}). Evaluating the relevant one-loop Feynman diagrams and constructing the dynamic RG flow equations give
$\chi_h=1/2$ and $z=3/2$ at the RG fixed point; the detailed calculations are well-document in the literature~\cite{fns,kpz} that we do not reproduce here. Since the disorder-induced nonlinear couplings in (\ref{eqdelrho}) are irrelevant in the long wavelength limit for $\rho_0\neq 1/2$, the remaining first order in space derivative linear term (\ref{eqdelrho}) can be removed by an appropriate Galilean boost, thereby reducing the governing equation for $\delta\rho$ to the $1d$ Burgers equation.

In the long wavelength limit, height fluctuations $h(x,t)$ follows  
the KPZ equation (\ref{kpz1}), for which the scaling of the time-dependent 
correlation function of $\delta \rho(x,t)$ is known: we have
\begin{equation}
 \overline {\langle [h(x,t)- h(0,0)]^2\rangle} =  |x|\Theta_h 
 (|x|^{3/2}/t), 
\end{equation}
corresponding to $\chi_\rho=-1/2$ and $z=3/2$~\cite{stanley,natter}; $\Theta_h$ is a scaling function. Therefore, the correlation function $C_{\rho\rho}(x,t)$ of the  density $\delta\rho(x,t)$
scales as $x^{-1}\Theta (|x|^{3/2}/t)$. 

 Further, there are no diverging fluctuation corrections to $\langle |\psi (k)|^2\rangle$, and hence it is still given by (\ref{lin-corr-psi-nothalf}) even in the full anharmonic theory in the long wavelength limit. This gives $\chi_\psi=-1/2=\chi_\rho$. Furthermore, the physical picture in the linearised theory that $\psi(x)$ can be obtained by minimising ${\cal F}_1$ still holds in the long wavelength limit.

We can then conclude that the correlation function of the total density $\rho(x,t)$
\begin{equation}
 C_{\rho}(x,t)\equiv\overline{\langle [\rho(x,t)-\rho(0,0)]^2\rangle} \label{crho}
\end{equation}
also scales as $x^{-1}\Theta (|x|^{3/2}/t)$,
same as what it would show in the absence of any quenched disorder. Thus disorder does not affect the universal scaling when the system is away from half-filling. In other words, experimental measurements of the scaling exponents in physical realisations of this model cannot detect if there is any disorder or not.


\subsection{Scaling near half-filling}\label{half-fill}

The physics near half-filling ($\rho_0\approx 1/2$) turns out to be very different  
from what we discussed above. In fact, the nonlinear terms involving quenched 
disorder turn out to be relevant, as we argue below. Notice first that the 
linear first order gradient term in space now vanish, i.e., $\lambda_{1\psi}=0$. Thus there are no underdamped 
propagating modes in the dynamics of $\delta\rho(x,t)$. For $\rho_0\approx 1/2$, as 
explained above, $\psi(x)$ is more relevant (in a scaling sense) than 
$\delta\rho(x,t)$. It is also more relevant than $\delta m(x)$ at linear level; see Eq.~(\ref{rcorr-psi-half}) above. This 
consideration allows us to write the equations for $\psi(x)$ and 
$\delta\rho(x,t)$, retaining only the most leading order nonlinear terms. We 
find
\begin{eqnarray}
 &&\nu_\psi\partial_{xx}\psi 
-\frac{\lambda_\psi}{2} \partial_x\psi^2 +\lambda_m\partial_x\delta m \nonumber\\&& -\lambda_{3\psi}\partial_x (\delta m\psi^2)=0, \label{psi-half}\\
&& \frac{\partial\delta \rho(x,t)}{\partial t} =\nu_{\rho}\partial_{xx} 
\delta \rho-\lambda_{\rho\psi}\partial_x (\psi\delta\rho)  + \partial_x f. 
\label{rho-half}
\end{eqnarray}
 We can extract a current $J_\rho$ from (\ref{rho-half}) above:
\begin{equation}
 J_\rho=-\nu_\rho\partial_x\delta\rho + \lambda_{\rho\psi} \psi\delta\rho - f,
\end{equation}
similar to the current $J$ in (\ref{burg-curr}) above. 
Notice that Eq.~(\ref{rho-half}) formally resembles the dynamical equation for a passive scalar, advected by a ``frozen-in'' Burgers-like irrotational velocity field~\cite{passive}. If we define a ``height field'' $h (x,t)$ via $\delta\rho(x,t)=\partial_x h(x,t)$, then in the long wavelength limit $h(x,t)$ satisfies
\begin{eqnarray}
 &&\frac{\partial h}{\partial t}=\nu_{\rho}\partial_{xx}h -\lambda_{\rho\psi} \psi \partial_x h  + f. \label{dis-kpz}
\end{eqnarray}
 While Eq.~(\ref{dis-kpz}) bears formal structural similarity with the pure $1d$ KPZ equation (\ref{kpz1}), there are fundamental differences. Unlike (\ref{kpz1}), Eq.~(\ref{dis-kpz}) is neither invariant under a Galilean transformation, nor it admits an FDT. Thus, unlike the $1d$ KPZ equation, not only the exponents $\chi_h$ and $z$ are {\em not} known exactly, there are {\em no} exact relation between them either.  In fact, diverging vertex corrections (see below) in (\ref{dis-kpz}) does not allow for a simple scaling relation between $\chi_h$ and $z$, unlike the KPZ equation. { This feature makes the problem at hand considerably more challenging than the pure $1d$ KPZ problem.

 {

Equation~(\ref{rho-half}), valid near half-filling,  may be considered as the hydrodynamic equation for the density of a collection of noninteracting particles moving on a rough but frozen surface, following the local slopes and under the effect of the annealed noise (since Eq.~(\ref{rho-half}) is linear in $\delta\rho$). Thus, close to the half-filled limit in the absence of any underdamped waves, a single ``$\delta\rho$'' particle is more likely to be found at ``valleys'' of the frozen surfaces than at the ``hills'', leading to roughness of the correlator of $\delta\rho$ roughly following the roughness of the frozen surface, modulated only by the diffusion by the nearest neighbour hopping and the annealed noise. This is expected to make $\chi_\rho> -1/2$ (the value of $\chi_\rho$ for a collection of noninteracting particles on a $1d$ smooth surface). We will see below that our detailed calculations confirm this intuitive argument.}
 
Before we embark on our pertubative dynamic RG calculations, we briefly refer to  an  exact mapping, via the Cole-Hopf transformation, between the $1d$ KPZ equation (\ref{kpz1}) and the equilibrium problem of a single directed polymer  with one transverse and one longitudinal directions in a random medium (DPRM)~\cite{kardar-prl,halpin-healy,kardar-book}.  It has been argued that in the ensuing competition between the elastic (kinetic) energy and the potential energy (due to the random potential) of the polymer, the latter dominates in $1d$. As a result, the polymer configuration is always {\em rough}, exact analogue of the rough phase of the $1d$ KPZ equation~\cite{halpin-healy,kardar-book}.} { The scaling exponents that characterise the fluctuations of the DPRM, are related to $\chi_h$ and $z$, and are calculated by using scaling arguments by balancing various terms in the energy contributions of the DPRM, as well as by more sophisticated functional RG methods~\cite{halpin-healy}. Unsurprisingly, this leads to the same scaling relation $\chi_h+z=2$, establishing the equivalence between the DPRM and the KPZ problem. In the present problem,  the presence of $\psi(x)$ (which is quenched) in Eq.~(\ref{rho-half}) makes any mapping to a DPRM-like problem less obvious. In fact, it can be shown that Eq.~(\ref{rho-half}) is formally equivalent to an imaginary time Schr\"odinger equation of a charged particle in a complex electromagnetic field; see Appendix \ref{cole-hopf}. Here, the presence of vertex corrections in (\ref{rho-half}) should preclude a simple scaling argument relating the scaling exponents. We do not explore this connection and its implications further here.}

As shown in Appendix \ref{half-filled}, the na\"ive perturbative fluctuation corrections to the 
model parameters due to nonlinear terms diverge in the long wavelength limit, 
necessitating systematic dynamic RG analysis. We confine ourselves to a low-order 
(one-loop) RG analysis, following the calculational scheme outlined above.

We provide the one-loop Feynman diagrams for the model parameters in Appendix \ref{half-filled}.  
We evaluate the integrals at the fixed dimension $d=1$, as done for 1D Burgers 
equation~\cite{stanley}.

The differential flow equations for the parameters are (using $d\tilde D/dl=0$)
\begin{eqnarray}
 &&\frac{d\nu_{\psi}}{dl}= \nu_{\psi}\left[-3+g\right],\\
 &&\frac{d\lambda_m}{dl}= \lambda_m \left[-\frac{5}{2}-\chi_\psi-2\bar g +\frac{g}{2}\right],\\
 &&\frac{d\lambda_{\psi}}{dl}=\lambda_{\psi}\left[-2+\chi_\psi-2g \right],\\
 &&\frac{d\lambda_{3\psi}}{dl}= \lambda_{3\psi}\left[-\frac{5}{2}+\chi_\psi \right],\\
 &&\frac{d\nu_{\rho}}{dl}=\nu_{\rho}\left[z-2+g_1 \right],\\
 &&\frac{dD}{dl}=D\left[-3+z-2\chi_\rho+2g_1\right],\\
 &&\frac{d\lambda_{\rho\psi}}{dl}=\lambda_{\rho\psi}\left[-1+z+ 
\chi_\psi-2g_1\right],
\end{eqnarray}
 Here, the effective coupling constants: $g=\frac{\tilde D \lambda^2_m 
\lambda^2_{\psi}}{\nu^4_{\psi} \varLambda^3},\, \bar 
g=\frac{\lambda_{3\psi}\lambda_m\tilde D}{\nu^2_{\psi}\varLambda},\, 
g_1=\frac{\lambda^2_{\rho\psi}\lambda^2_m \tilde D}{\nu^2_{\psi} 
\nu^2_{\rho}\varLambda^3}$.  The flow equations for the coupling constants are:
 \begin{eqnarray}
  &&\frac{dg}{dl}= g\left[ -7g -4\bar g +3 \right],\label{flowg}\\
  &&\frac{d\bar g}{dl}=\bar g [-\frac{3g}{2}-2\bar g +1 ],\label{flow-gbar}\\
  &&\frac{dg_1}{dl}=g_1 \left[ -6g_1 -4\bar g -g+3 \right].\label{flowg1}
 \end{eqnarray}
 Coupling constants $g$ and $\bar g$ have their origins in (\ref{psi-half}), where as $g_1$ appears in the dynamics of $\delta\rho$. Unsurprisingly, (\ref{flowg}) and (\ref{flow-gbar}) do not depend upon $g_1$. At the RG fixed point, $dg/dl=0=d\bar g/dl=dg_1/dl$.
 Flow equations (\ref{flowg}) and (\ref{flow-gbar}) admit several solutions at the fixed point $(g^*,\,\bar g^*)$: (i) (0,0), (ii) (3/7,\,0), (iii) (0,\,1/2) and (iv) (1/4,\, 5/16). Out of these, $(g^*,\,\bar g^*)=(1/4,\,5/16)$ is the globally stable fixed point. The RG flow around the fixed points are shown in Fig.~\ref{rg-flow}.
 \begin{figure}[htb]
  \includegraphics[width=7cm]{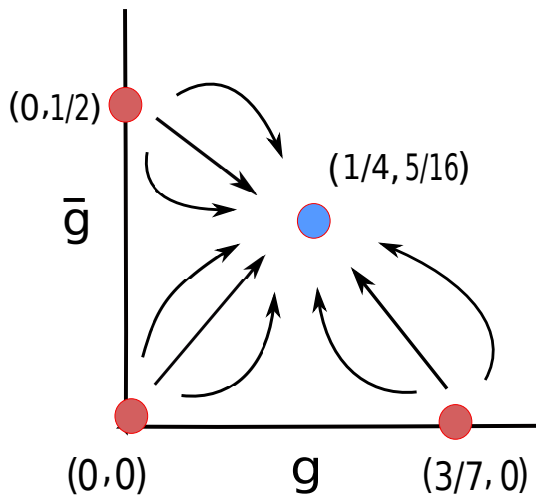}
  \caption{Fixed points and RG flow lines in the $g-\overline{g}$ plane. The small blue circle with
  $(g^*,\,\bar g^*)=(1/4,\,5/16)$ is the fully stable fixed point; arrows indicate the flow directions in the $g-\bar g$ plane.}\label{rg-flow}
 \end{figure}

  By setting $dg_1/dl=0$, we find $g_1^*=1/4$ at the stable fixed point. This gives
  \begin{eqnarray}
  \chi_\psi&=&-\frac{1}{4}=-0.25,\\
  \chi_\rho&=&-\frac{3}{8}=-0.375,\\
  z&=&2-\frac{1}{4}=\frac{7}{4}=1.75.\label{scaling-expo}
  \end{eqnarray}
  by using the above RG flow equations.  Now the scaling exponent $\chi_h$ of $\delta\rho(x,t)=\partial_x h(x,t)$ is given by $\chi_h=\chi_\rho+1= 5/8=0.625$. { These exponents imply $\beta=5/14$~\cite{imply}.} Notice also that $\chi_h+z=19/8$, which is larger than 2,  in contrast to the KPZ equation (see above). Thus, the scaling exponents belong to a {\em new} universality class that is distinct from the $1d$ KPZ universality class.   Furthermore, these exponents also imply that both $\overline{\psi(x)^2}$ and $\overline{\langle \delta\rho(x,t)^2 \rangle}$ are {\em finite} in the thermodynamic limit, as they should be for a number conserving system.  This also implies that linear theory-based picture that $\psi(x)$ can be obtained by minimising ${\cal F}_2$ may still hold in the long wavelength limit if we construct ${\cal F}_2$ in terms of the {\em renormalised parameters}. Clearly the values of the scaling exponents (\ref{scaling-expo}) are larger than their counterparts for the pure $1d$ KPZ equation, in agreement with the trends observed in Refs.~\cite{mustansir-prl,stinchcombe}.   
  { The renormalised time-dependent height fluctuation correlation function $C_{hh}(x,t)$ then reads
\begin{equation}
 C_{hh}(x,t)=|x|^{2\chi_h}g_h(x^z/t)=|x|^{1.25}g_h(x^{1.75}/t),\label{final-scale}
\end{equation}
where $g_h$ is a dimensionless scaling function. The scaling exponents $\chi_h$ and $z$ may conveniently calculated by measuring  the width $W_h(t)$ defined as
\begin{equation}
 W_h(t)=\sqrt{\langle[h(x,t)-h(x,0)-\overline h(t)]^2\rangle},
\end{equation}
where $\overline h(t)$ is the mean height at time $t$. In a system of size $L$, $W_h(t)$ scales as $t^{\chi_h/z}$ for small $t$, and $ L^{\chi_h}$ for large $t$; see Refs.~\cite{mustansir-prl,stanley}.}

{ 

\section{New universality class: How close must $\rho_0$ be to 1/2?}\label{heu}

In Sec.~\ref{half-fill} above, we have argued that when $\rho_0$ is ``close'' to 1/2, the quenched disorder becomes relevant, and a new universality class emerges. The moot question then is, how close must $\rho_0$ be to 1/2 in the thermodynamic limit in order for the new universality to emerge? Numerical results of Refs.~\cite{mustansir-pre,mustansir-prl} indicate that there is actually a (small) window around $\rho_0=1/2$ within which new universal behaviour is expected to be observed. We now argue that this result could be understood within our RG calculations on the hydrodynamic equations. 

It is evident that the irrelevance of the quenched disorder in our calculations for $\rho_0\neq 1/2$ essentially originates from the linear first order space derivative terms in (\ref{psieq}) and (\ref{eqdelrho}) to dominate over the diffusive terms. Assuming that our continuum hydrodynamic equations are valid up to a small scale $2\pi/\Lambda$, corresponding to an upper wavevector cutoff $\Lambda$, there could be a situation where for sufficiently high wavevectors (but still well-within the validity of our hydrodynamic equations), the diffusive terms dominate. For concreteness, let us first focus on (\ref{psieq}) and assume $\nu_\psi q^2 \gg \lambda_{1\psi} q$ for $\Lambda > q> q_\psi$. As we start eliminating modes from $q=\Lambda$, our calculational scheme as outlined in Sec.~\ref{half-fill} holds, $\nu_\psi$ starts getting positive corrections, where as $\lambda_{1\psi}$ gets none. Nonetheless, as the procedure of mode eliminating persists, $\lambda_{1\psi} q$ eventually dominates over $\nu_\psi q^2$, at some wavevector around $q_\psi$. There are then two possiblities: (i) the scale separation between $\Lambda$ and and $q_\psi$ is large, so that $\nu_\psi$ (and the system) gets enough ``renormalisation group time'' $l$, such that fluctuation corrections can change the scaling of $\nu_\psi q^2$ to $\nu_\psi q^{7/4}$, or (ii) the scale separation is small, fluctuation corrections cannot change the scaling of $\nu_\psi q^2$ as in the linearised version of (\ref{psieq}). Since beyond this scale, $\lambda_{1\psi}$ dominates, any further mode elimination at wavevectors $q<q_\psi$ produces only finite corrections to $\nu_\psi$ and $\lambda_{1\psi}$, leaving scaling around the crossover scale unchanged all the way to $q\rightarrow 0$. Thus, even if $\rho_0\neq 1/2$, it is possible to have the scaling of $\psi$ getting renormalised by the disorder, if there is a sufficient scale separation as explained above (or, equivalently, a non-zero but sufficiently small $\lambda_{1\psi}$; see below).  

Similar analysis holds for Eq.~(\ref{eqdelrho}) as well. That is, here too, even when $\lambda_{1\rho}$ is non-zero, two possible scenarios can emerge: (i) if there is a large enough scale separation between $\Lambda$ and $q_\rho$, below which $\lambda_{1\rho}$ dominates over $\nu_\rho q^2$, fluctuation corrections renormalise $\nu_\rho q^2$ to $\nu_\rho q^z$ with $z=7/4$, or, (ii) if the scale separation is not large enough, fluctuation corrections then disorder is not adequate to renormalise $\nu_\rho q^2$. As mode elimination proceeds below the crossover scale $q_\rho$, $\lambda_{1\rho}$ dominates and no further relevant corrections to $\nu_\rho$ is obtained, if scenario (i) prevails, we have $z=7/4$ (see above), or if scenario (ii) emerges, then disorder is irrelevant, and the KPZ universality class (as for the pure system) holds.

We now make an estimate of the window around $\rho_0=1/2$ within which non-KPZ, disorder-dependent new universal behaviour is expected to be found. In this part of our analysis, for simplicity, we assume that all the nonlinear coupling constants in (\ref{psieq}) and (\ref{eqdelrho}) are ${\cal O}(1)$, ignore their distinctions and represent all of them by the single notation $\lambda_e$; we also set $\nu_\psi\sim\nu_\rho\sim \nu,\,\lambda_{1\psi}\sim \lambda_{1\rho} \sim c$. Now, according to the logic outlined above, in order for the non-KPZ disorder-controlled universal behaviour to be observed, we must first have a crossover from $\nu k^2$ dominated regime to a regime where $\nu k^2$ gets renormalised to $\nu_R k^z$ due to the fluctuation corrections {\em before} $ck$ dominates as one moves from high-$k$ to low-$k$; here $\nu_R$ is the amplitude of renormalised $\nu$ and scales with $\tilde D$. This allows us to set a threshold $c_0$ for $c$:
\begin{equation}
 c_0k\sim\nu_Rk^z\sim \nu k^2,
\end{equation}
for some crossover scale $k$. This yields the crossover scale $k$ and $c_0$ in terms of the other model parameters. For $c<c_0$, or $\rho_0<1/2 + c_0/2$, disorder is relevant and new universal behaviour emerges; else, for $c>c_0$, or $\rho_0 >1/2 + c_0/2$, disorder is irrelevant and KPZ universality class would ensue. Particle-hole symmetry of the model then tells us that within the range $\rho_0=1/2 \pm c_0/2$, disorder is relevant; outside this window disorder is irrelevant. Further, since in our perturbative calculations, $\nu_R\propto \tilde D$, the extent of this window given by $c_0$ should increase with $\tilde D$, indicating a disorder distribution dependent window. Our conclusions in this Section are in general agreement with those in Ref.~\cite{mustansir-pre}, where they indeed found the extent of such a window to depend upon the parameters that specify the disorder distribution. }

\section{Summary and outlook}\label{summ}

We have thus shown how the $1d$ KPZ equation with columnar disorder emerges as the hydrodynamic theory for the number conserving asymmetric exclusion processes in the presence of short range random quenched disorder.  This not invariant under a Galilean transformation, nor it satisfies an FDT, unlike the pure $1d$ KPZ equation.  By using this continuum hydrodynamic theory, we show that when the system is away from half-filled,  the density fluctuations propagate across the system in the form of traveling waves. As we have argued above, this leads to the density fluctuations encountering only an average effect of  the quenched disorder, rendering it  irrelevant (in a RG sense), and restoring the Galilean invariance and FDT in the effective long wavelength theory. As a result, the local density fluctuations display spatio-temporal scaling belonging to the $1d$ KPZ universality class in the long wavelength limit. Thus experimental measurements of the scaling of the density fluctuations cannot detect any quenched disorder.  In contrast, when the system is close to being half-filled, the propagating modes are absent and the density fluctuations are  overdamped. Thus, in the absence of any ``averaging effects'' due to the travelling waves, the density fluctuations are strongly affected by the presence of the quenched disorder, resulting in a new universality class. The Galilean invariance and FDT are broken in the effective long wavelength theory, making the universality class distinctly different from its $1d$ KPZ counterpart. The spatial scaling exponent $\chi_\rho$ (and hence $\chi_h$) and the dynamic exponent $z$ are larger when the system is near the half-filled limit than when it is away. Equivalently, the height field is rougher and has a slower relaxation near the half-filled limit than when it is away.  These features are corroborated in the numerical simulations of the related agent-based model~\cite{mustansir-prl,stinchcombe}. However. unlike the $1d$ KPZ equation,  the scaling exponents of $\delta \rho$ are not known exactly, due to the lack of invariance of Eq.~(\ref{rho-half}) under a Galilean transformation and the absence of an FDT.  } 

{  Numerical results of Refs.~\cite{mustansir-pre,mustansir-prl} have shown the existence of domain walls in the density snapshots in the regimes where the quenched disorder is relevant. Since our density correlations are averaged over the disorder configurations, and the domain wall can appear anywhere in the system depending upon the particular disorder configuration,  the density correlations calculated here do not reveal their existence. Nonetheless, a larger roughness exponent for relevant quenched disorder provides indirect evidence for bigger hills and valleys in a given density snapshot, that may mimic domain walls. On the whole, our hydrodynamic theory successfully classifies two possible universal behaviour in number conserving TASEP with short ranged quenched disordered hopping rates, in agreement with the numerical results~\cite{mustansir-prl,stinchcombe}.}

   We have argued in Appendix \ref{high-d} that even in higher dimensions, the model should belong to the KPZ universality class (for that dimension) when away from half-filling, but new universal behaviour is expected close to half-filling.  We hope our analytical results will give strong impetus for further detailed numerical studies on lattice-gas models or on the hydrodynamic equations developed here (see, e.g., Ref.~~\cite{erwin-jstat}) in the future that will complement our analytical studies here.

These results complement the recent studies on number conserving TASEP with quenched but non-random hopping rates~\cite{lebo,niladri1,tirtha-prr}, where  two generic types of steady states was found, that revealed the existence of a type of universality. Similarly, in the present study too, the scaling of the density fluctuations are very sensitive to whether the system is away or close to half-filling. It would be interesting to explore any deeper connections between our studies and  these models further.

{ Our hydrodynamic theory should pave the way to construct hydrodynamic theories for similar but more complex quenched disordered driven systems with number conservation, e.g., disordered TASEP in closed networks~\cite{rakesh-tasep}.}
In this work, we have confined ourselves to study the effects of short ranged quenched disorder. Long range quenched disorder is expected to be relevant and further modify the scaling exponents obtained here. In fact, for sufficiently long range random quenched disorder, the scaling exponents are likely to be affected {\em even when the system is away from half-filling}. However, even then the present analysis suggests that the scaling exponents near the half-filling will be different from their counterparts away from half-filling. A full quantitative analysis of long range random quench disorder will be presented elsewhere.

\section{Acknowledgement}

The authors thank J. Toner for helpful discussions in the early stage of this work, and M. Barma for a critical reading of the manuscript, and U. T\"auber for constructive suggections and helpful comments.

\appendix

\section{Derivation of the hydrodynamic equations}\label{derivation}
\subsection{Asymmetric case}
Consider a closed 1D lattice with $L$ sites with $a$ as the lattice spacing. Hence, the total length of the lattice is $La$. Let $\overline N=L/a$. We are interested in the thermodynamic limit, $\overline N\rightarrow\infty$, which can be realised, e.g., for $L\rightarrow \infty$ for a fixed $a$, or vice versa.

{ We closely follow Ref.~\cite{kiran} in deriving the hydrodynamic limit.}
We start by restating the dynamical equations for the occupation number $n_i$ at site $i$ of the lattice:
\begin{equation}
 \frac{\partial n_i(t)}{\partial t}= m_{i-1} n_{i-1} (1-n_{i}) - m_{i} n_{i} (1-n_{i+1}),\label{disc1}
\end{equation}
where $m_i$ is the hopping rate from site $i$ to $i+1$. Total particle number $N=\sum_i n_i$ is clearly a constant of motion. 
We define $x=i/L$, which becomes quasi-continuous in the thermodynamic limit. Further, suitable coarse-graining allows us to obtain { a continuous density $\rho(x)=\langle n_i\rangle_c$  and $\rho(x-1/L)=\langle n_{i-1}\rangle_c$}, where as $\langle...\rangle_c$ implies a suitable coarse-graining.  Neglecting correlations between neighbouring sites (in the spirit of a mean-field like approach),  we obtain 
\begin{equation}
 \frac{\partial\rho(x)}{\partial t} = m(x-\frac{1}{L})\rho(x-\frac{1}{L})[1-\rho(x)] - m(x) \rho(x) [1-\rho(x+\frac{1}{L})].
\end{equation}
In the hydrodynamic limit, we expand $\rho(x-1/L),\,\rho(x+1/L)$ and $m(x-1/L)$ up to second order in $1/L$. This { together with a ballistic rescaling of time} gives (\ref{fulrho}) in the main text (where a conserved noise has been added). We further identify $\nu=m_0/L$ as the effective diffusion coefficient.  Since microscopic diffusivity $\nu\sim 1/L$, it gets vanishingly small in the thermodynamic limit, allowing the nonlinear effects to dominate over a wide range of length scales; see Ref.~\cite{dhruba} for a related context.  For a periodic system as here, the Fourier wavevectors are labelled as $2n\pi/L$, where $n=0,\,\pm 1,\,\pm 2,..$ In the limit of large $L$, $k$ becomes quasi-continuous starting from $k\rightarrow 0$. 

\subsection{Symmetric case}
Let us now consider the symmetric case, where particles can move bidirectionally subject to exclusions. The equation motion for occupation $n_i$ at site $i$ is given by
\begin{eqnarray}
 \frac{\partial n_i}{\partial t}&=& m_{i-1}n_{i-1}(1-n_i) - m_{i-1}n_i (1-n_{i-1}) \nonumber \\&-& m_i n_i (1-n_{i+1}+m_i n_{i+1} (1-n_i).
\end{eqnarray}
As before, coarse-graining and expanding up to ${\cal O}(1/L^2)$, we obtain
\begin{equation}
 \frac{\partial \rho}{\partial t}=\frac{1}{L^2} \frac{\partial}{\partial x}\left(m(x)\frac{\partial \rho}{\partial x}\right).
\end{equation}
This gives, upon discarding irrelevant nonlinearities and rescaling $t$ by $1/L^2$ (diffusive time-scale),
\begin{equation}
 \frac{\partial \rho}{\partial t}=m_0 \frac{\partial^2\rho }{\partial x^2} +\partial_x f,
\end{equation}
where a conserved noise $\partial_x f$ has been added; this is identical to the coupled equations (\ref{sep-psi}) and (\ref{sep-rho}) in the main text.

\section{Generating functional}\label{functional}
The generating functional~\cite{jansen} corresponding to Eqs.~(\ref{psieq}) and (\ref{eqdelrho}) is 
\begin{equation}
 {\mathcal Z}= \int {\mathcal D}\psi {\mathcal D}\hat{\psi} {\mathcal D}\delta\rho {\mathcal D}\hat{\rho} {\mathcal D} \delta m \exp\{-{\mathcal S}\}.
\end{equation}
Here, $\hat\psi(x)$ and $\hat\rho(x,t)$ are the dynamic conjugate fields to $\psi(x)$ and $\delta\rho(x,t)$, respectively~\cite{jansen}. Further, $\cal S$ is the action functional given by
\begin{eqnarray}
 {\mathcal S}&=& \int dx~ \hat{\psi} [ -\nu_{\psi} \partial_{xx}\psi -\lambda_{1\psi}\partial_x \psi - \lambda_m \partial_x \delta m + \frac{\lambda_{\psi}}{2}\partial_x \psi^2\nonumber \\
 && - \lambda_{2\psi}\partial_x(\delta m \psi) + \lambda_{3\psi}\partial_x(\delta m \psi^2)] + \int dx\frac{\delta m^2}{4 \tilde D} \nonumber\\
 &+& \int dx dt~  D(\partial_{xx}) \hat{\rho} \hat{\rho} + \hat{\rho} [ \partial_t \delta\rho -\lambda_{1\rho}\partial_x \delta\rho -\nu_{\rho} \partial_{xx}\delta\rho \nonumber \\
 &&+\frac{\lambda_{\rho}}{2}\partial_x{\delta\rho^2} - \lambda_{2\rho}\partial_x(\delta m \delta\rho) + \lambda_{\rho\psi}\partial_x(\delta\rho \psi) \nonumber\\
 &&+ \lambda_{3\rho}\partial_x(\delta m \delta\rho^2) + \lambda_{3\rho\psi}\partial_x(\delta m \psi \delta\rho) ].\label{action1}
\end{eqnarray}

The two-point autocorrelation functions of the fields in the harmonic theory, i.e., neglecting all the nonlinear terms are
\begin{eqnarray}
  &&\overline{ |\psi_{{ k},\omega}|^2 } = \frac{2\tilde{D}\lambda^2_m \delta(\omega)}{\lambda^2_{1\psi}+\nu^2_{\psi} k^2}\approx \frac{2\tilde{D}\lambda^2_m \delta(\omega)}{\lambda^2_{1\psi}},\label{psi-nothalf-corr} \\
  &&\overline{\langle |\delta\rho_{{ k},\omega}|^2 \rangle} = \frac{2Dk^2}{(\omega+\lambda_{1\rho}{ k})^2+\nu^2_{\rho} k^4},\label{rho-nothalf-corr},\\
  &&\overline{\langle |h_{{ k},\omega}|^2 \rangle}= \frac{2D}{(\omega+\lambda_{1\rho}{ k})^2+\nu^2_{\rho} k^4} \label{h-nothalf-corr};
  \end{eqnarray}

We set up the perturbation theory by expanding in the coefficients of the nonlinear terms. The nonlinear vertices are represented diagrammatically as given in Fig.~\ref{bare-vert}.
\begin{figure}[h]
\includegraphics[width=7cm]{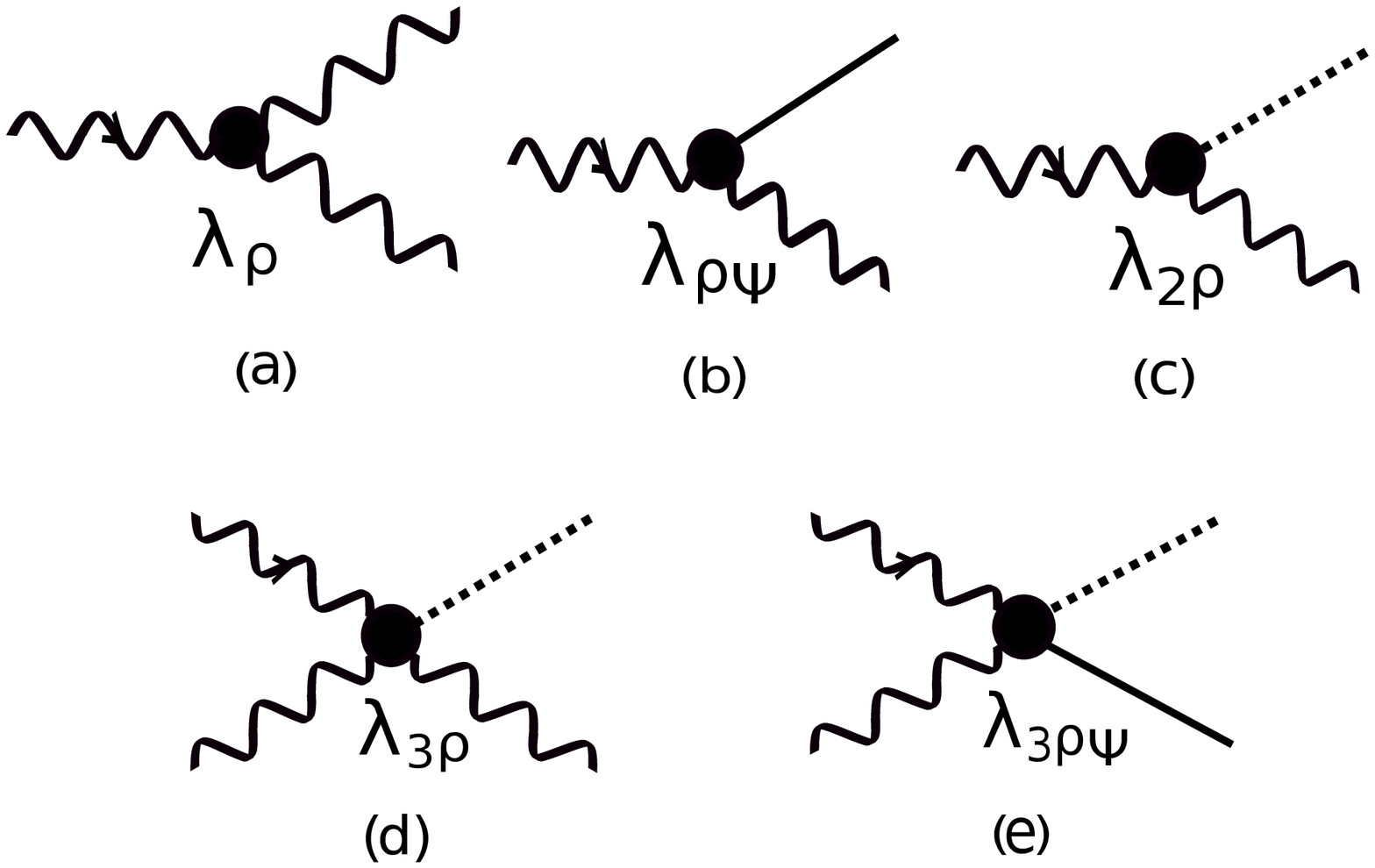}
\includegraphics[width=7cm]{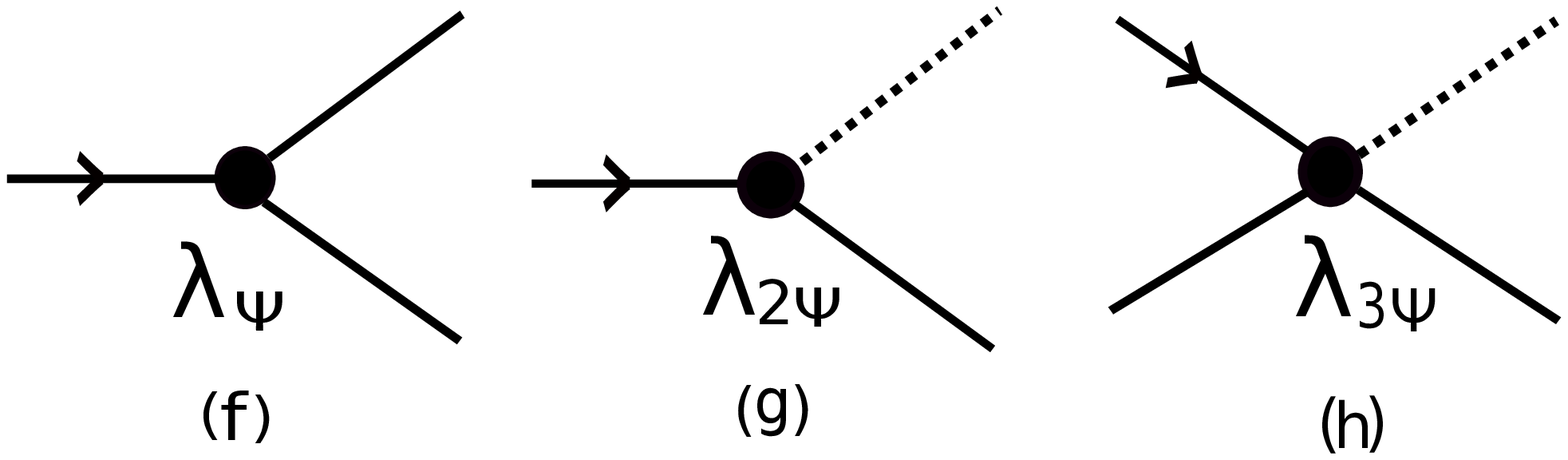}
\caption{Diagrammatic representations of the nonlinear vertices in action (\ref{action1}).}\label{bare-vert}
\end{figure}

The bare or unrenormalised propagators and the correlators are diagrammatically represented as given in Fig.~\ref{bare-corr}.
\begin{figure}[h]
 \includegraphics[width=7cm]{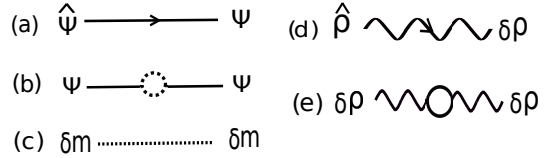}
 \caption{Diagrammatic representations of the lines representing the propagators and correlators of $\rho$, $\psi$ and $\delta m$.}\label{bare-corr}
\end{figure}

\section{Perturbation theory for $\rho_0\neq 1/2$}\label{not half-filled}
 \subsection{One-loop corrections}
The one-loop diagrams are constructed out of the propagators and correlators of $\rho$, $\psi$ and $\delta m$, which are diagrammatically represented as follows:

\begin{figure}[h]
 \includegraphics[width=7cm]{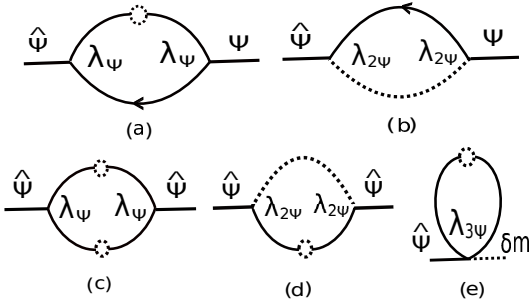}
 \caption{One-loop corrections to $\nu_\psi$,  and $\lambda_m$, originated from nonlinear vertices of action (\ref{action1}).}\label{psi-corrections}
\end{figure}
One loop corrections for parameters in (\ref{psieq}):
\begin{eqnarray}
  &&{\rm Fig.}~(\ref{psi-corrections}a)= \frac{2\lambda_\psi^2 \lambda_m^2 \tilde D}{\lambda_{1\psi}^3} [ik] \int \frac{dq}{2\pi} .\\
  && {\rm Fig.}~(\ref{psi-corrections}b)= \frac{2\lambda_{2\psi}^2 \tilde D}{i\lambda_{1\psi}} [ik] \int \frac{dq}{2\pi} .\\
&& {\rm Fig.}~(\ref{psi-corrections}c)=  \left[\frac{\tilde D \lambda_\psi \lambda_m^2}{\lambda_{1\psi}^2} \right]^2 [k^2]\int \frac{dq}{2\pi} .\\
&& {\rm Fig.}~(\ref{psi-corrections}d)= 2\left[\frac{\tilde D \lambda_m \lambda_{2\psi}}{\lambda_{1\psi}}\right]^2[k^2] \int \frac{dq}{2\pi} .\\
&& {\rm Fig.}~(\ref{psi-corrections}e)= -\frac{\lambda_{3\psi} \tilde D \lambda_m^2}{\lambda_{1\psi}^2}[ik] \int \frac{dq}{2\pi}.
\end{eqnarray}

The one-loop corrections for the propagator and auto-correlator for $\rho$ that in turn contribute to the fluctuation corrections of $\nu_\rho$ and $D$ are shown below:
\begin{figure}[h]
 \includegraphics[width=7cm]{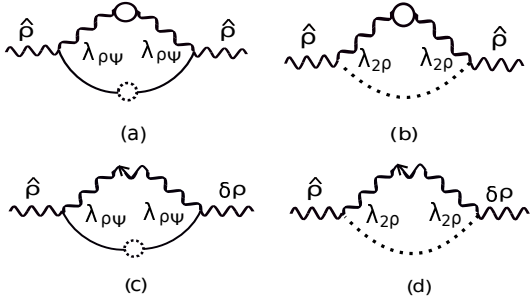}
 \caption{One-loop corrections to $\nu_\rho$ and $D$, originated from the disorder-dependent nonlinear vertices of action (\ref{action1}).}\label{rho-corrections}
\end{figure}

One loop corrections for parameters in (\ref{eqdelrho}) are
\begin{eqnarray}
 && {\rm Fig.}~(\ref{rho-corrections}a)=  \frac{\lambda_{\rho\psi}^2 D \tilde D \lambda_m^2}{\lambda_{1\psi}^2 \lambda_{1\rho}} [k^2] \int \frac{dq}{2\pi}  .\\
 && {\rm Fig.}~(\ref{rho-corrections}b)=   \frac{2D\tilde D \lambda_{2\rho}^2}{\lambda_{1\psi}^2 \lambda_{1\rho}}  [ k^2] \int \frac{dq}{2\pi} .\\
 && {\rm Fig.}~(\ref{rho-corrections}c)= -\frac{2\tilde D \lambda_{\rho\psi}^2  \lambda_m^2}{\lambda_{1\psi}^2 \lambda_{1\rho}} [ik] \int \frac{dq}{2\pi},\\
 && {\rm Fig.}~(\ref{rho-corrections}d)= -\frac{2\tilde D \lambda_{2\rho}^2}{ \lambda_{1\rho}} [ik] \int \frac{dq}{2\pi}.
\end{eqnarray}

{ In addition, there are diverging fluctuation corrections to these parameters from the nonlinear vertices which survive in the pure (no disorder) limit. These are standard~\cite{stanley} and we do not evaluate them here.}

The perturbative step of evaluating the one-loop diagram is followed by $q\rightarrow bq,\; \omega \rightarrow b^z \omega$. Together with the rescaling of space(or momentum) and time(or frequency), long wavelength parts of the fields are rescaled as follows:
\begin{eqnarray}
&&\hat\psi(q)=b^{-\chi_\psi-1}\hat\psi(bq),\;\psi=b^{1+\chi_\psi}\psi(bq),\nonumber \\&&\delta m(q)=b^{1/2}\delta m(bq),\\ 
&&\hat \rho(q,\omega)=b^{z-\chi_\rho}\hat \rho(bq,b^z\omega),\nonumber \\&&\delta\rho(q,\omega)=b^{1+z+\chi_\rho}\delta\rho(bq,b^z\omega). \label{rescaling11}
\end{eqnarray}


\subsection{Scaling of the coupling constants near the Gaussian fixed point}\label{appen-gauss}

We now determine the scaling of the dimensionless effective coupling constants that depend upon $\tilde D$, the disorder variance, near the Gaussian fixed point. Away from the half-filling, these 
 are  $a_0=\frac{\lambda_\psi^2 \tilde D \lambda_m^2}{\lambda_{1\psi}^4}$, $a_1=\frac{\tilde D \lambda_{2\psi}^2}{\lambda_{1\psi}^2}$, $a_2=\frac{\tilde D \lambda_m \lambda_{3\psi}}{\lambda_{1\psi}^2}$, $a_3=\frac{\lambda_{\rho\psi}^2 \tilde D \lambda_m^2}{\lambda_{1\rho}^2\lambda_{1\psi}^2}$ , $a_4=\frac{\lambda_{2\theta}^2\tilde D}{\lambda_{1\rho}^2}$. Under the rescaling of the momentum, frequency and the fields as defined above, we find near the Gaussian fixed point
 \begin{eqnarray}
  &&\frac{da_0}{dl}= -a_0,\, \frac{da_1}{dl}= -a_1,\, \frac{da_2}{dl}= -a_2,\nonumber\\
  &&\frac{da_3}{dl}= -a_3,\, \frac{da_4}{dl}= -a_4.
 \end{eqnarray}
Thus all these coupling constants are irrelevant at $1d$ in the RG sense near the Gaussian fixed point.
Unsurprisingly, the dimensionless coupling constant $\tilde g=\frac{\lambda_\rho^2 D}{\nu_\rho^3}$ that appears in the pure KPZ problem is {\em relevant} near the Gaussian fixed point at $1d$:
\begin{equation}
 \frac{d\tilde g}{dl}= \tilde g .
\end{equation}
Thus the Gaussian fixed point that controls the scaling of the linear theory is stable with respect to perturbations by the quenched disorder when away from half-filling, but unstable with respect to the pure KPZ anharmonic effects at $1d$, leading to the universal scaling belonging to the $1d$ KPZ universality class when away from half-filling.

The situation changes drastically close to half-filling. The relevant dimensionless coupling constants $g,\,\overline g$ and $g_1$ are all relevant near the Gaussian fixed point, as can be seen from the flow equations (\ref{flowg}-\ref{flowg1}) above. Thus, the linear theory scaling should be affected by the quenched disorder; see next Section for details.

\section{Perturbation theory for $\rho_0\approx 1/2$: Dynamic RG analysis}\label{half-filled}

The action functional now reads
\begin{eqnarray}
 {\mathcal S}=&& \int dx~ \hat{\psi} [ -\nu_{\psi} \partial_{xx}\psi - \lambda_m \partial_x \delta m + \frac{\lambda_{\psi}}{2}\partial_x \psi^2 \nonumber\\
 && + \lambda_{3\psi}\partial_x(\delta m \psi^2)] + \frac{\delta m^2}{4 \tilde D}  + \int dx dt~ D(\partial_{xx}) \hat{\rho} \hat{\rho}\nonumber\\
 && + \hat{\rho} \left[ \partial_t \delta\rho -\nu_{\rho} \partial_{xx}\delta\rho  + \lambda_{\rho\psi}\partial_x(\delta\rho \psi) \right] .\label{action2}
\end{eqnarray}
The autocorrelation functions of the fields in harmonic theory for the action (\ref{action2}) are given by
 \begin{eqnarray}
  &&\overline{ |\psi_{{ k},\omega}|^2 } = \frac{2\tilde{D}\lambda^2_m \delta(\omega)}{\nu^2_{\psi} k^2},\label{psi-half-corr} \\
  &&\overline{\langle |\delta\rho_{{ k},\omega}|^2 \rangle} = \frac{2Dk^2}{\omega^2+\nu^2_{\rho} k^4},\label{rho-half-corr}\\
  &&\overline{\langle |h_{{ k},\omega}|^2 \rangle} = \frac{2D}{\omega^2+\nu^2_{\rho} k^4}.\label{h-half-corr}
\end{eqnarray}
We highlight  a technical point here. With (\ref{rcorr-psi-half}), at half-filling
\begin{equation}
 \overline{[\psi(x)]^2}=\frac{2\tilde D\lambda_m^2}{\nu^2_\psi}\int_{1/L}^{1/a}\frac{dk}{2\pi} \frac{1}{k^2}
\end{equation}
grows linearly with $L$ without any bound. On the other hand, $\overline{\langle\delta \rho(x,t)^2\rangle}$ remains finite in the thermodynamic limit for any filling-factor and $\overline{[\psi(x)]^2}$ remains finite in the thermodynamic limit when the system is away from half-filling. That $\overline{[\psi(x)]^2}$ diverges in the thermodynamic limit is not consistent with the particle number conservation in the system. We shall see below that nonlinear effects make $\overline{[\psi(x)]^2}$ finite in the thermodynamic limit. 

For $\rho_0\approx 1/2$, some of the nonlinear coefficients in the action functional (\ref{action1}) vanish, giving rise to the action (\ref{action2});  hence some of the one-loop diagrams that exist for $\rho_0\neq 1/2$ now vanish.

One loop corrections for parameters in (\ref{psi-half}), (\ref{rho-half}) and hence in  near the half-filled limit action (\ref{action2}) are

\begin{eqnarray}
 && {\rm Fig.~} (\ref{psi-corrections}a)= \frac{\tilde D \lambda_m^2 \lambda_\psi^2}{\nu_\psi^3}[-k^2] \int \frac{dq}{2\pi} ~ \frac{1}{q^4} ,\\
  &&  {\rm Fig.~}(\ref{psi-corrections}c)= \frac{\tilde D^2 \lambda_m^4 \lambda_\psi^2}{\nu_\psi^4}[k^2] \int \frac{dq}{2\pi} ~ \frac{1}{q^4} ,\\
  && {\rm Fig.~}(\ref{psi-corrections}e)= \frac{2\tilde D \lambda_m^2 \lambda_{3\psi}}{\nu_\psi^2}[-ik] \int \frac{dq}{2\pi} ~ \frac{1}{q^2} ,\\
 && {\rm Fig.~}(\ref{rho-corrections}a)= \frac{2\lambda_{\rho\psi}^2 D\tilde D \lambda_m^2}{\nu_\psi^2 \nu_\rho^2} [k^2]\int \frac{dq}{2\pi}  ~ \frac{1}{q^4} ,\\
  &&  {\rm Fig.~}(\ref{rho-corrections}c)= \frac{\lambda_{\rho\psi}^2 \tilde D \lambda_m^2}{\nu_\psi^2 \nu_\rho} [-k^2]\int \frac{dq}{2\pi}  ~ \frac{1}{q^4} .
\end{eqnarray}

{Note that {\rm Fig.~}(\ref{psi-corrections}a), {\rm Fig.~}(\ref{psi-corrections}c) contribute to $\psi$ the propagator and correlator of $\psi$, respectively, and {\rm Fig.~}(\ref{rho-corrections}a), {\rm Fig.~}(\ref{rho-corrections}c) contribute to the correlator and propagator of $\delta\rho$, respectively  for $\rho_0\approx 1/2$.}

\begin{figure}[h]
 \includegraphics[width=7cm]{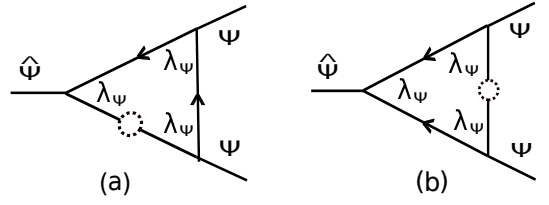}
 \caption{One-loop vertex correction diagrams for $\lambda_\psi$ when $\rho_0\approx 1/2$.}\label{psi-vertex-corrections}
 \end{figure}
 
\begin{figure}[h]
 \includegraphics[width=4cm]{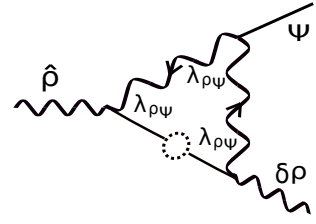}
 \caption{ One-loop vertex correction diagram for $\lambda_{\rho\psi}$ when $\rho_0\approx 1/2$. In addition to this diagram, there are two other diagrams which survive in the pure limit of the problem, and sum of which vanish in the long wavelength limit.}\label{rho-vertex-corrections}
\end{figure}
One-loop contributions for parameters of nonlinearity in (\ref{action2}) are
\begin{eqnarray}
  &&{\rm Fig.~}(\ref{psi-vertex-corrections}a)= \frac{2\tilde D \lambda_m^2 \lambda_{\psi}^3}{\nu_\psi^4}[ik] \int \frac{dq}{2\pi} ~ \frac{1}{q^4},\\
  &&  {\rm Fig.~}(\ref{psi-vertex-corrections}b)= \frac{\tilde D \lambda_m^2 \lambda_{\psi}^3}{\nu_\psi^4}[-ik] \int \frac{dq}{2\pi} ~ \frac{1}{q^4},\\
  &&{\rm Fig.~}(\ref{rho-vertex-corrections})= \frac{2\lambda_{\rho\psi}^3 \tilde D \lambda_m^2}{\nu_\psi^2 \nu_\rho^2} [ik]\int \frac{dq}{2\pi}  ~ \frac{1}{q^4} .
\end{eqnarray}

Once the fields having support in the wavevector range $\Lambda/b$ to $\Lambda$ are integrated out ($b=\exp(\delta l)>1$), we obtain ``new'' model parameters corresponding to a modified action ${\cal S}^<$ having $\Lambda/b$ as the wavevector upper cutoff. We obtain
\begin{eqnarray}
 &&\nu_{\psi}^<=\nu_{\psi}\left[ 1+\frac{\tilde D \lambda_m^2 \lambda_{\psi}^2}{\nu_{\psi}^4 } \int_{\varLambda/b}^{\varLambda} \frac{dq}{2\pi} \frac{1}{q^4} \right],\nonumber\\
 &&\lambda_m^<=\lambda_m [1-\frac{2\lambda_{3\psi}\lambda_m \tilde D }{\nu_{\psi}^2}\int_{\varLambda/b}^{\varLambda}\frac{dq}{2\pi} \frac{1}{q^2} \nonumber \\
 &&~~~~~~~~~~~~~~~ +  \frac{\tilde D \lambda_m^2 \lambda_{\psi}^2}{2\nu_{\psi}^4 } \int_{\varLambda/b}^{\varLambda} \frac{dq}{2\pi} \frac{1}{q^4} ],\nonumber\\
 &&\lambda_{\psi}^<= \lambda_{\psi} \left[ 1- \frac{2\tilde D \lambda_m^2 \lambda_{\psi}^2}{\nu_{\psi}^4 } \int_{\varLambda/b}^{\varLambda} \frac{dq}{2\pi} \frac{1}{q^4} \right],\nonumber\\
 &&\nu_{\rho}^<=\nu_{\rho}\left[ 1+\frac{\tilde D \lambda_m^2 \lambda_{\rho\psi}^2}{\nu_{\psi}^2 \nu_{\rho}^2} \int_{\varLambda/b}^{\varLambda} \frac{dq}{2\pi} \frac{1}{q^4} \right],\nonumber\\
 && D^<=D\left[ 1+ \frac{2\tilde D \lambda_m^2 \lambda_{\rho\psi}^2}{\nu_{\psi}^2 \nu_{\rho}^2} \int_{\varLambda/b}^{\varLambda} \frac{dq}{2\pi} \frac{1}{q^4} \right],\nonumber\\
 && \lambda_{\rho\psi}^<=\lambda_{\rho\psi} \left[ 1-\frac{2\tilde D \lambda_m^2 \lambda_{\rho\psi}^2}{\nu_{\psi}^2 \nu_{\theta}^2} \int_{\varLambda/b}^{\varLambda} \frac{dq}{2\pi} \frac{1}{q^4} \right].
\end{eqnarray}

\section{Generation of higher order nonlinear terms}

For $\rho_0\approx 1/2$, $\chi_\psi>0$ in the linear theory. This implies na\"ive perturbation expansion should generate higher order nonlinear terms in $\psi$ that will remain relevant since $\chi_\psi>0$. In our above dynamic RG analysis, we have neglected such perturbatively generated nonlinearities. In the renormalised theory, $\chi_\psi<0$ rendering all such perturbatively generated higher order nonlinear terms irrelevant near the RG fixed point. Technically speaking, if $g_h$ is an effective coupling constant that has its origin from one such higher order nonlinear term, then in the renormalised theory near the RG fixed point, writing schematically,
\begin{equation}
 \frac{dg_h}{dl} =  \Delta_g g_h + {\cal O} (g_h)^2,\label{high-drg}
\end{equation}
where $\Delta_g <0$ is the scaling dimension of $g_h$ near the stable RG fixed point obtained above, evaluated using renormalised $\chi_\psi<0$. Hence, for sufficiently small (bare) $g_h$, it flows to zero in the long wavelength limit. For larger values of bare $g_h$, it is possible that $g_h$ grows in the long wavelength limit near the RG fixed point, depending upon the signs of the possible higher order terms in (\ref{high-drg}). Since in the present theory, contributions from such perturbatively generated higher nonlinear terms to the model parameters should essentially scale with higher powers of $\tilde D$, these will be important for large $\tilde D$, for which positivity of $m(x)$ is not guaranteed and we do not expect our theory to be valid in that regime. \\

\section{Cole-Hopf transformation}\label{cole-hopf}

We apply the Cole-Hopf transformation to the  dynamical equation for $\delta \rho(x,t)$ near the half-filled limit. We start from (\ref{rho-half}) and also retain the standard KPZ nonlinear term in $\delta \rho (x,t)$ for calculational reasons:
\begin{equation}
 \frac{\partial h}{\partial t}=\nu_{\rho}\partial_{xx}h -\lambda_{\rho\psi} \psi \partial_x h -\frac{\lambda_\rho}{2}(\partial_x h)^2 + f.\label{new-delrho}
\end{equation}
Now define the Cole-Hopf transformation~\cite{halpin-healy,stanley,kardar-book} $h(x,t)\equiv (2\nu/\lambda_\rho) \ln W$, and apply  it on (\ref{new-delrho}) above to get
\begin{equation}
 \frac{\partial W}{\partial t}=\nu\partial_{xx} W -\lambda_{\rho\psi} \psi \partial_x W + \frac{\lambda_\rho}{2\nu} fW.\label{cole-eq}
\end{equation}
If $\lambda_{\rho\psi}=0$, then (\ref{cole-eq}) is the equation for a partition function of a single DP in a time-dependent random potential, or the imaginary time Schr\"odinger equation for  a particle in a 
time-dependent random potential~\cite{halpin-healy,kardar-book}. When $\lambda_{\rho\psi}\neq 0$, (\ref{cole-eq}) may be written in a slightly different form as
\begin{widetext}
\begin{equation}
 \frac{\partial W}{\partial t}=\nu\left(\partial_x -\frac{\lambda_{\rho\psi}}{\nu}\psi(x)\right)^2 W +\lambda_{\rho\psi}(\partial_x \psi) W- \frac{\lambda^2_{\rho\psi}}{\nu}\psi^2 (x) W(x,t) + \frac{\lambda_\rho}{2\nu} f W(x,t). \label{schro-eq}
\end{equation}
\end{widetext}
Equation~(\ref{schro-eq}) may be interpreted as the imaginary time Schr\"odinger equation for a charged particle in a random quenched vector potential $A(x)\equiv i\frac{\lambda_{\rho\psi}}{\nu}\psi(x)$  and a random electrostatic potential that has a quenched piece  $\frac{\lambda^2_{\rho\psi}}{\nu}\psi^2 (x)- \lambda_{\rho\psi}\partial_x\psi(x)$ and a time-dependent random piece $\frac{\lambda_\rho}{2\nu}f(x,t)$. This has the potential of making hitherto unexplored connections with other areas of condensed matter physics or statistical mechanics; see, e.g., Ref.~\cite{hatano}. This could be studied in the future.

\section{Generalisation to higher dimensions}\label{high-d}

 We now briefly comment on the nature of scaling and critical dimensions of the model after generalising to $d$-dimensions. At dimensions $d>1$, when away from half-filling the propagating modes make the system anisotropic. We investigate the relevance of the coupling constants $a_0,~ a_1,~ a_2,~ a_3,~ a_4,~ \tilde g$ near the Gaussian fixed point, when away from half-filling. We find
\begin{eqnarray}
 &&\frac{da_0}{dl}= -da_0,\, \frac{da_1}{dl}= -da_1,\, \frac{da_2}{dl}= -da_2,\nonumber\\
  &&\frac{da_3}{dl}= -da_3,\, \frac{da_4}{dl}= -da_4.\\
 &&\frac{d\tilde g}{dl}= \tilde g [2-d].
\end{eqnarray}
Thus, the pure KPZ nonlinear coupling $\tilde g$ is the most relevant coupling constant. Hence, $\tilde g$ should control the roughening transition (smooth-to-rough transition) and $d=2$ should be the (lower) critical dimension as in the pure KPZ equation~\cite{natter}; further, spatial isotropy should be restored at large scales. This led us to speculate that the universal scaling in the rough phase of this model, not accessible perturbatively, along with the upper critical dimension, should be identical to that in the pure KPZ equation. The latter for the pure KPZ problem is speculated to be four by some studies, although it is still debatable~\cite{ucd-kpz,ucd-kpz1}.

Close to half-filling, the situation changes dramatically. We generalise the action (\ref{action2}) to $d$-dimensions by simply considering the space coordinate $\bf x$ to be $d$-dimensional. Close to 1/2-filling, the underdamped propagating waves vanish and the model is fully isotropic. The appropriate coupling constants $g,\overline g$ and $g_1$ near the Gaussian fixed point follow (obtained by suitably generalising the rescaling (\ref{rescaling11}) above)
\begin{eqnarray}
 \frac{dg}{dl}&=& g[4-d] ,\nonumber\\
 \frac{d\bar g}{dl}&=& \bar g[2-d] ,\nonumber\\
 \frac{dg_1}{dl}&=& g_1[4-d].
\end{eqnarray}
These na\"ively suggest that the lower critical dimension might be higher than two (might actually be 4). These may be systematically investigated by using, e.g., the Cole-Hopf transformation discussed above~\cite{erwin-epl}. If so, then the perturbatively inaccessible rough phase and the corresponding upper critical dimension should be different from those in the pure KPZ problem.  The latter is likely to be at least as hard to infer as it is for the pure KPZ problem. Nonetheless, if the lower critical dimension is indeed raised to 4, the corresponding upper critical dimension is likely to be higher than 4. A more complete discussion will be made available elsewhere in the future.


\begin{thebibliography}{99}
\bibitem{foot1} We implicitly impose $\delta m(x)< m_0$, so that the local hopping rate $m(x)$ remains positive everywhere, a condition needed for finite currents in the steady states of asymmetric exclusion processes.
\bibitem{imply} There are some quantitative discrepancies between the scaling exponents (\ref{scaling-expo}) and those reported in the earlier numerical studies. These could be due to the low order nature of our perturbative calculations for the scaling exponents; better systematic numerical studies would be useful in this regard, although getting good quality data from a large enough system with quenched disorder is admittedly a difficult task. Our statements on the universality classes are independent of the numerical accuracy of the scaling exponents, and agree with Refs.~\cite{mustansir-prl,stinchcombe}.
\end{thebibliography}
\end{document}